\documentclass[prb,aps,twocolumn,floatfix,superscriptaddress,showpacs]{revtex4-1}
\pdfoutput=1
\usepackage{ifpdf}
\usepackage{hyperref}
\usepackage{amsmath,amssymb}
\usepackage[utf8]{inputenc}
\usepackage{graphicx}
\usepackage{color}
\renewcommand{\vec}[1]{\mathbf{#1}}
\newcommand{\sech}{\mathop{\mathrm{sech}}}

\begin{document}
\title{Lindblad theory of dynamical decoherence of quantum-dot
  excitons}

\author{P.~R.~Eastham}
\affiliation{School of Physics, Trinity College, Dublin 2, Ireland.}  

\author{A.~O.~Spracklen}
\affiliation{SUPA, School of Physics and Astronomy, University of
  St. Andrews, KY16 9SS, U.K.}

\author{J.~Keeling}
\affiliation{SUPA, School of Physics and Astronomy, University of
  St. Andrews, KY16 9SS, U.K.}

\begin{abstract}
  We use the Bloch-Redfield-Wangsness theory to calculate the effects
  of acoustic phonons in coherent control experiments, where
  quantum-dot excitons are driven by shaped laser pulses. This theory
  yields a generalized Lindblad equation for the density operator of
  the dot, with time-dependent damping and decoherence due to phonon
  transitions between the instantaneous dressed states. It captures
  similar physics to the form recently applied to Rabi oscillation
  experiments [A. J. Ramsay \textit{et al.},  Phys. Rev. Lett.  104,
  017402 (2010)], but guarantees positivity of the density
  operator. At sufficiently low temperatures, it gives results
  equivalent to those of fully non-Markovian approaches [S. L\"uker
  \textit{et al.},  Phys. Rev. B 85, 121302 (2012)], but is
  significantly simpler to simulate. Several applications of this
  theory are discussed. We apply it to adiabatic rapid passage
  experiments, and show how the pulses can be shaped to maximize the
  probability of creating a single exciton using a frequency-swept
  laser pulse. We also use this theory to propose and analyze methods
  to determine the phonon density of states experimentally, i.e.
  phonon spectroscopy, by exploring the dependence of the effective
  damping rates on the driving field.
\end{abstract}


\maketitle

\newcommand{\tgunkv}[1]{{\vec{#1},t}}
\newcommand{\tgunk}{\tgunkv{k}}

\section{Introduction}

Controlled manipulation of coherent quantum systems is a crucial
requirement for quantum information technologies, can be exploited in
ultrafast switches, and may allow the exploration of exotic regimes of
quantum dynamics. An important example among solid-state systems is
that of excitons in quantum dots,\cite{Vamivakas2010} which provide
discrete atomic-like transitions that can be manipulated using optical
pulses. These transitions have been demonstrated experimentally to
correspond to two-level systems, for which resonant optical excitation
induces Rabi oscillations.\cite{Stievater2001,Ramsay2010b} Thus under
pulsed excitation the number of excitons created oscillates with the
pulse area: a pulse of the correct duration and intensity creates
exactly one exciton in the quantum dot, while other pulses create
superpositions of one- and no- exciton states. Frequency-swept laser
pulses have also been used to create single excitons in quantum dots,
implementing the protocol of adiabatic rapid passage (ARP), and
allowing state manipulation in a way which is robust against
fluctuations in the coupling strengths and transition energies of the
dots.\cite{Wu2011,Simon2011,Schmidgall2010,Eastham2009}

A theoretical description of such coherent control experiments must
capture both the dynamics of the driven quantum dot, and the
scattering and decoherence introduced by the interaction between the
dot and its environment. In particular, the coupling to acoustic
phonons leads to dephasing of the Rabi oscillations\cite{Ramsay2010a}
and limits the inversion in ARP.
\cite{Wu2011,Simon2011,Luker2012,Reiter2012}  (Optic phonons could
play a role in ARP with very intense, very short
pulses,\cite{Schuh2011} where the energy scales are comparable to the
optic phonon energies, but we do not consider this regime here.) The
standard theoretical approach\cite{Breuer2007} involves second-order
perturbation theory in the phonon coupling, leading to an equation of
motion for the reduced density operator of the dot. This equation
involves an integral over all previous states of the dot, capturing
the memory effects due to the finite bandwidth and response time of
the environment.

A frequently used approach to treating this type of non-Markovian
equation of motion is a form of Markov approximation which reduces the
equation to a time-local equation, with the effects of the phonons
appearing as a constant Lindblad form describing dephasing.  Such an
equation would be valid under the assumption that the response time of
the environment is the shortest time scale in the problem. This
approximation, however, is generally incorrect for quantum dot
excitons.\cite{Wilson-Rae2002,Vagov2007,Ramsay2010} It predicts that
the environment induces transitions independently of the state of the
dot, in contradiction to experiments in which the driving field
changes the dephasing. Such dynamical excitation-dependent dephasing
has, however, been successfully described by more sophisticated
approaches. These include numerically exact path integral
methods\cite{Vagov2007,McCutcheon2011,Luker2012,Reiter2012}
(quasi-adiabatic propagator path integral, QUAPI); systematic
expansions in exciton-phonon
correlations~\cite{Forstner2003,Luker2012,Reiter2012} up to fourth
order and including memory effects; and time-local
approximations\cite{Ramsay2010a,Ramsay2010,Ramsay2011,McCutcheon2010,Nazir2008}
allowing for some of the memory effects neglected in the simplest form
of Born-Markov approximation.  These time-local approximations improve
on the simplest Born-Markov approximation by calculating the decay
rates arising from system-bath coupling making use of the actual
system Hamiltonian, including driving.  This same point, of using the
actual system Hamiltonian rather than a non-interacting approximation,
is also crucial in describing the correct equilibrium state of
strongly coupled systems, as discussed in
Ref.~\onlinecite{Cresser1992}.

When there is strong coupling to phonons but weak driving of the dot,
accurate results can be found by making a polaron
transformation\cite{McCutcheon2010}, so that coupling to phonons
appears in the driving term; one may then treat the driving term in
the Born approximation, and derive effective excitation/de-excitation
rates depending on the detuning of the drive, and the phonon density
of states, accounting for multi-phonon excitations.  Such an approach
can also be extended to cavity-QED situations\cite{Wilson-Rae2002},
with the driving replaced by a coupling to a cavity.  It has also been used
to study the fluorescence of a driven quantum dot, coupled to a
cavity.\cite{Roy2011,Roy2012,Roy2012a,Hughes}  However, the treatment
of the driving within the Born approximation limits the validity of
this approach to weak driving.  Alternatively, one may view this as the
statement that at strong driving, the dot state changes too rapidly
for the phonons to follow, and so the polaron picture breaks down.
This has been extensively studied recently
\cite{McCutcheon2010,McCutcheon2011} using a variational polaron
transformation~\cite{Silbey1984}.  The variational approach can reproduce
the exact results of the path integral across a range of driving
strengths.  In the limit of strong driving and experimentally relevant
dot-phonon couplings, \citet{McCutcheon2011} show that the time-local
approaches discussed
above~\cite{Ramsay2010a,Ramsay2010,Ramsay2011,McCutcheon2010,Nazir2008}
become increasingly accurate, as strong driving breaks the polaron
picture, and relatively weak dot-phonon coupling (at low temperatures)
allows both a Markovian approximation and the neglect of multiphonon
effects.

In this paper, we consider situations where a dot is strongly driven,
and with experimentally relevant (i.e., relatively weak) dot-phonon
couplings.  We discuss the application of the
Bloch-Redfield-Wangsness\cite{Wangsness1953,Redfield1955} (BRW)
theory, widely used to describe nuclear spin relaxation, to coherent
control experiments on excitons in quantum dots. It is similar to the
simplest Born-Markov approximation, but allows the time scale set by
the inverse level spacing of the Hamiltonian to be smaller than the
response time of the environment. We use this theory to derive a
generalized Lindblad form for the phonon-induced damping, in which the
transition operators connect the time-dependent eigenstates of the
dot, with the expected perturbative transition rates
[see Eq.~(\ref{eq:4})]. This differs from the simplest Lindblad form
mentioned above, which is often applied to quantum dots, in which the
transition operators do not necessarily connect
eigenstates. Furthermore, in contrast to the forms obtained by some
other time-local approximations,\cite{Ramsay2010,Ramsay2011} it
guarantees the positivity of the density operator, and so can be used
across a wider variety of pulse shapes and temperatures.  We focus on
the application to determining the effects of phonons for adiabatic
rapid passage in quantum dots, and show that the results are similar
to those recently obtained from the correlation expansion at fourth
order.\cite{Luker2012,Reiter2012} That method includes all memory
effects of the environment and allows for some phonon correlations,
and is known to be
accurate\cite{Luker2012,Reiter2012,Glassl2011,Vagov2011} for the
parameters relevant here (since it agrees with the exact path-integral
results). Thus Eq.~(\ref{eq:4}) provides a simple picture of the
effects of dephasing and a lightweight computational approach for
modeling dephasing in quantum dots. As a further application of this
theory, we use it to demonstrate the feasibility of measuring the
phonon spectra and distribution functions, by exploiting the
driving-field dependence of the effective damping rates. We propose
and analyze two forms of such spectroscopy, one based on a
generalization of the ARP experiment, and one based on the response to
off-resonant continuous-wave excitation.

These Markovian approximations are appropriate for strong driving; at
weak driving (a limit which is avoided in the remainder of this paper),
differences arise from the exact solution of the independent boson
model (IBM). For infinitesimal driving, the independent boson model
can be analytically solved~\cite{Duke1965,Mahan2000} by finding the
linear absorption spectrum about the undriven state: $A(\omega)
\propto \int dt e^{-i \omega t} \exp[ \varphi(t) - \varphi(0) ],
\varphi(t) = \int d \omega e^{i \omega t} n_B(\omega)
J(\omega)/\omega^2$. For the realistic phonon spectral function
$J(\omega)$ considered in this paper [see Eq.~(\ref{eq:modelspecdens})],
the exact absorption spectrum consists of an unbroadened
zero-phonon-line (ZPL), and sidebands associated with one or many
phonon events.  At $4$K (the temperature considered for pulse
optimization below), the ZPL contains 86\% of the spectral weight.  If
the Markovian approach we use is applied in the limit of weak driving
(outside the range of validity as discussed above), it predicts only
the ZPL, and misses the small phonon sidebands. The origin of this
discrepancy (at $4$K) is, however, not a consequence of multiphonon
effects [one may safely expand the expression for $A(\omega)$ to
linear order in $J(\omega)$], but of the Markov approximation.  The
origin of this discrepancy is as discussed in
Refs.~\onlinecite{Ford1996,Ford1999,Lax2000,Ford2000}: Markovian
approaches sample the bath at a frequency dependent on the system
Hamiltonian, while absorption spectra depend on the bath response at
the probe frequency.  In the limit of vanishing driving, the Markovian
approximation, as discussed below, produces no linewidth, hence it
matches only the dominant ZPL part of the exact solution.  For any
non-zero driving (i.e. beyond linear response), the Markovian approach
produces a non-zero linewidth, however there is no analytic solution
of the IBM for finite driving.  Thus, to test our theory in this
regime requires comparison to numerical approaches; such a comparison
to existing~\cite{Luker2012} numerical results is, indeed, given below,
and the match is seen to be very good (better than the match for
vanishing driving). That is as to be expected, given the central point
of recent
work~\cite{Ramsay2010,Ramsay2010a,McCutcheon2010,McCutcheon2011} on
``excitation induced dephasing'': the presence of strong driving
strongly affects the effective dephasing and dissipation rates, and
the behavior at vanishing driving does not control how the system
responds at strong driving.

The remainder of this paper is structured as follows. In Sec.\
\ref{sec:eom}, we outline the derivation of the equation of motion for
a driven quantum dot interacting with acoustic phonons. In Sec.\
\ref{sec:arpinv}, we present the predictions of this equation for the
inversion (exciton occupation) obtained in ARP, discuss how this
process may be optimized in the presence of phonon-induced dephasing,
and explain why the dephasing can in some circumstances improve the
final inversion. In Sec.\ \ref{sec:phononspec}, we outline the
application to phonon spectroscopy.  In Sec.\ \ref{sec:discuss}, we
discuss further the relationship between the positivity-preserving
Lindblad form [Eq.~(\ref{eq:4})], obtained here, and the generally
positivity-violating precursor to this form [Eq.~(\ref{eq:2})], which
is sometimes used directly.\cite{Ramsay2010,Ramsay2011} We present
numerical results showing the failure of this latter approximation in
the case of ARP pulses. In Sec.\ \ref{sec:conc}, we summarize our
conclusions. Finally, the Appendix provides details of the derivation
of Eq.~(\ref{eq:4}).

\section{Equations of motion}
\label{sec:eom}

In this section, we present the derivation of the secularized
(Lindblad) form of the equation of motion for the reduced density
matrix $\rho(t)$. Our approach initially follows the same steps as in
Refs.~\onlinecite{Ramsay2010,Ramsay2011}.  However, in order to
produce an approach that preserves positivity of the density matrix
throughout the range of validity of perturbation theory, we must
additionally secularize the resulting
equations.\cite{Dmcke1979,Spohn1980} For completeness, we include in
this paper also the steps which follow
Refs.~\onlinecite{Ramsay2010,Ramsay2011}.  In this section, we outline
the main steps of the derivation, and present further details in
the Appendix.

We consider a single quantum dot, driven close to one of its discrete
transition frequencies by a laser pulse with a time-dependent
amplitude and frequency. For simplicity we assume that the excitation
is circularly polarized, so that only one of the exciton spin states
is relevant and we may neglect the formation of biexcitons. Thus we
model the dot as a two-level system, which may be in the ground state,
$|0\rangle$, or the one-exciton state, $|X\rangle$. The Hamiltonian
may be expressed, using pseudospin-1/2 operators $\vec{s}$, as
($\hbar=1$) \begin{equation}H_{\mathrm{dot}}=\Delta(t)s_z - \Omega(t)
  s_x, \label{eq:ham}\end{equation} in the rotating-wave
approximation, and in a frame rotating at the instantaneous driving
frequency $\omega(t)$. Here $\Delta(t)=E_0-\omega(t)$ is the detuning
between the transition energy $E_0$ and the instantaneous driving
frequency. $\Omega(t)$ is the time-dependent Rabi frequency,
corresponding to the amplitude of the driving
pulse. $\vec{s}=\vec{\sigma}/2$, where the Pauli operator
$\sigma_z=|X\rangle\langle X|-|0\rangle\langle 0|$ describes the
occupation of the exciton state, while $\sigma_x=|X\rangle\langle
0|+|0\rangle\langle X|=s_++s_- $ describes the electric dipole moment
of the transition. We refer to $s_z=\sigma_z/2$ as the inversion.

We focus on the effects of acoustic phonons, which are the dominant
dephasing mechanism in recent Rabi
flopping~\cite{Ramsay2010,Ramsay2011} and ARP
experiments.\cite{Wu2011,Simon2011,Luker2012,Reiter2012} They couple to the dot
via the deformation potential
coupling, 
\begin{equation}H_{\mathrm{c}}=s_z \sum_q (g_q b_q+g_q^\ast
  b_q^\dagger),
  \label{eq:1}
\end{equation} where $q$ labels the phonon wavevectors,
$b_q$ ($b_q^\dagger$) is a phonon annihilation (creation) operator,
and $g_q$ is the coupling constant. The phonon effects are
controlled by the phonon spectral density, $J(\omega)=\sum_q
|g_q|^2\delta(\omega-\omega_q)$. We take the model used in Ref. \onlinecite{Ramsay2010}
for a GaAs/InGaAs quantum dot, \begin{equation}
  J(\omega)=\frac{\hbar A}{\pi k_B}\omega^3
  e^{-\omega^2/\omega_c^2},\label{eq:modelspecdens} \end{equation}
with similar parameters $A=11.2 \mathrm{\ fs\ K^{-1}}$, $\hbar\omega_c=2$~meV. In
Eq.~(\ref{eq:modelspecdens}), the low-frequency behavior
$J(\omega)\propto \omega^3$ arises from the coupling and
density-of-states for acoustic phonons, while the high-frequency
cut-off at $\omega_c$ arises from the size of the dot; confined
excitons do not couple effectively to phonons of wavelengths smaller
than the confinement.

\begin{figure}[htpb]
  \centering
  \ifpdf
  \includegraphics[width=3.2in]{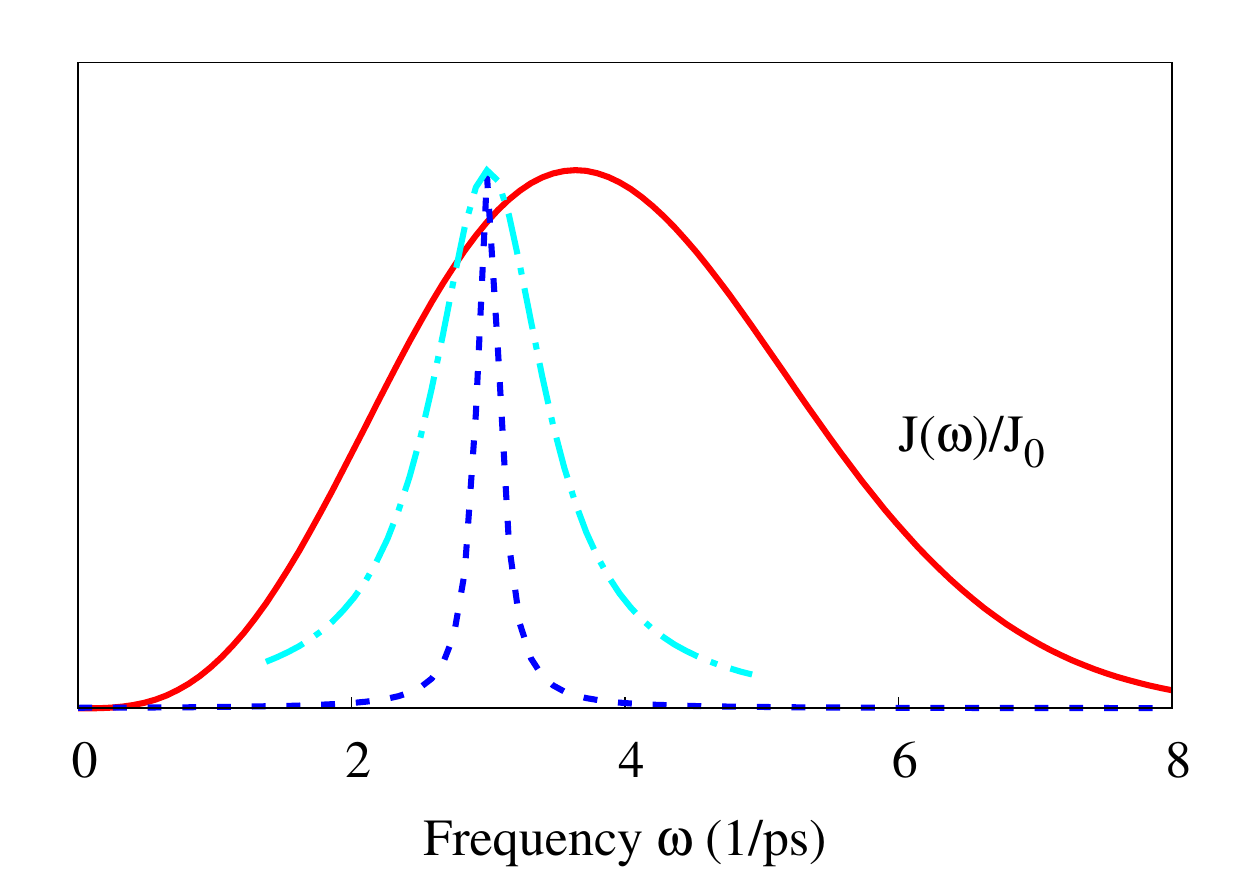}
  \else
  \includegraphics[width=3.2in]{cartoon.eps}
  \fi
  \caption{(color online). Illustration of time scales required for validity of
    (time-dependent) Markov approximation, shown in the frequency
    domain.  Solid red line shows the frequency dependent decay rate,
    illustrated by the phonon spectral function $J(\omega)$. At low
    temperatures and slow driving (dashed line), the linewidth is
    sufficiently small that the decay rate does not vary significantly
    across the linewidth (bath correlation time is short compared to
    decay time).  At higher temperatures, or faster driving, the
    linewidth grows (dot-dashed line) so that the decay rate does
    vary (decay time or sweep time become comparable to bath
    correlation time). The lineshapes illustrated are Lorentzians
    corresponding to transition rates $T_1^{-1}=0.2$ and 1.0
    $\mathrm{ps^{-1}}$, which may be compared with
    Fig.~\ref{fig:dampfig}.}
  \label{fig:timescales}
\end{figure}

In the limit that $\Delta(t)$ and $\Omega(t)$ vary slowly with time, we may
treat them effectively as constants and use the approach discussed in
Refs.~\onlinecite{Ramsay2010,Ramsay2011}, so that the effect of the
acoustic phonons can be found by transforming to the interaction
picture, and using the Born-Markov approximation there. This requires
that the interaction picture density operator $\tilde\rho(t)$ is
approximately constant over the correlation time of the phonon bath
($\sim 1/\omega_c$), so that $\tilde{\rho}(t^\prime) \simeq
\tilde{\rho}(t)$ on the right-hand side of
Eq.~(\ref{eq:mastereq}). Equivalently, this means that the bath
density of states should be flat over the effective linewidth of the
system, as illustrated in Fig.~\ref{fig:timescales}. Such an
approximation is valid as long as neither the decay rate nor sweep
rates ($\dot\Delta/\Delta,\dot\Omega/\Omega$) are too high, as both
contribute to the effective linewidth. Note that applying the
Born-Markov approximation directly in the Schr\"odinger picture
requires additionally that the density of states is flat on the scale
set by the position of the line, i.e., the energy scale of the
Hamiltonian, and this is not the case here.

It is convenient to introduce rotated spin operators
$\vec{r}=R\vec{s}$, with $R$ a rotation by angle $\tan^{-1}
\Omega/\Delta$ around the y-axis, so that the instantaneous system
Hamiltonian becomes $H_{\mathrm{dot}} = \Lambda r_z$ where $\Lambda =
\sqrt{\Omega^2 + \Delta^2}$ is the dressed-state splitting.  For the
acoustic phonon coupling considered here, this yields (see the
Appendix for further details):
\begin{equation}
  \label{eq:2}
  \dot{\tilde{\rho}} 
  = - P Q \tilde{\rho} +  Q \tilde{\rho} P
  + 
  P \tilde{\rho} Q^\dagger  - \tilde{\rho} Q^\dagger P,
\end{equation}
where $P, Q$ are time-dependent operators of the  form:
\begin{align}
  \label{eq:3}
  P(t) &=
  \frac{\Delta}{\Lambda} r_z +
  \frac{\Omega}{2\Lambda}(r_+e^{i\Lambda t}+r_- e^{-i\Lambda t}),
  \\
  \label{eq:10}
  Q(t) &= \int d \nu J(\nu)\!
  \int^t\!\!\! dt^\prime P(t^\prime)
  \nonumber\\&\qquad\times
  \left[(n_\nu +1) e^{-i \nu (t-t^\prime)} + n_\nu e^{i \nu (t-t^\prime)} \right],
\end{align} and $n_{\nu}$ is the phonon occupation function at
frequency $\nu$.  After undoing the transformation to the interaction
picture, this gives the density matrix equation form corresponding to
the results in Ref.~\onlinecite{Ramsay2011}.  

This equation is not of Lindblad form, and consequently it can lead to
density matrix evolution that violates positivity.  For the
relatively short pulses in Ref.~\onlinecite{Ramsay2010,Ramsay2011},
one may readily check that this is not a problem.  However, for our
application to ARP pulses, positivity violation can occur at late
times under conditions where the perturbative approximations required
for Born-Markov remain valid; this is discussed further in
Sec.~\ref{sec:discuss}

\begin{figure}
  \ifpdf
  \includegraphics[width=3.3in]{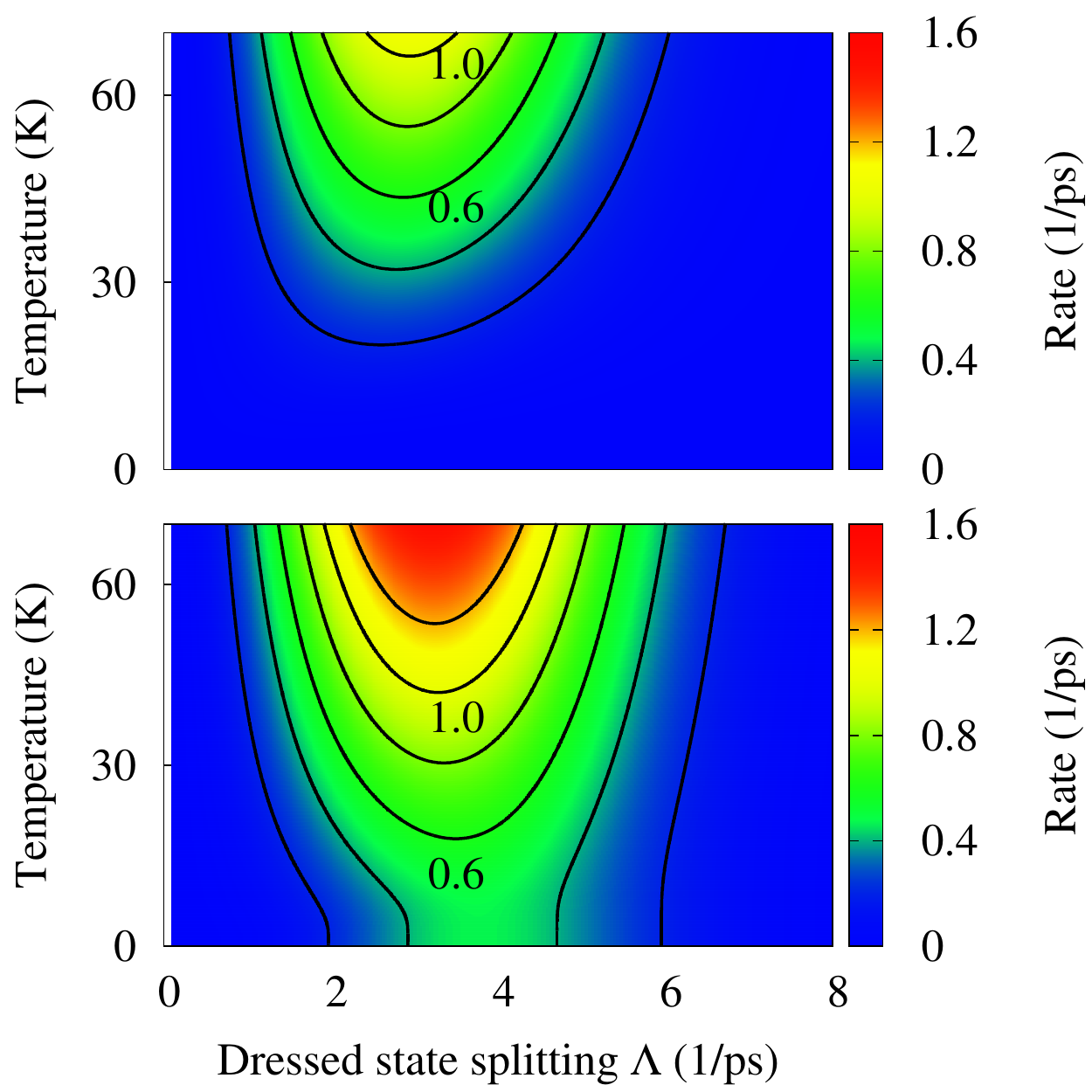}
  \else
  \includegraphics[width=3.3in]{dampfig.eps}
  \fi
  \caption{(color online). Phonon absorption (top) and emission
    (bottom) rates, Eqs. (\ref{eq:absrate}) and (\ref{eq:emrate}),
    respectively, divided by the squared ratio of Rabi splitting to
    dressed-state energy splitting, $\Omega^2/\Lambda^2$, as a
    function of the dressed-state energy splitting
    $\Lambda=\sqrt{\Omega^2+\Delta^2}$; for resonant driving
    $\Delta=0$ and the scaling is one.}
\label{fig:dampfig}
\end{figure}

This issue of positivity violation was discussed extensively in, e.g.,
Ref.~\onlinecite{Dmcke1979}, where it was shown that there exists more
than one form of Markovian density matrix equation which faithfully
represents the infinitesimal increment of the full density matrix
evolution in the Markovian (perturbative) limit.  However, although
these different forms are equivalent regarding infinitesimal
timesteps, those equations that are not of Lindblad form do not
conserve positivity.  A Lindblad form can nonetheless be derived by
averaging so as to remove the rapidly oscillating terms in
Eq.~(\ref{eq:2}) --- such a procedure, known as secularization, yields
a Lindblad form that in the perturbative limit is equally valid to
Eq.~(\ref{eq:2}). Further details are presented in the Appendix. After
transforming back to the Schr\"odinger picture, the result is
\begin{equation} 
  \begin{aligned}
    \dot{\rho} = &-[\gamma_{a}(\Omega,\Delta)/2] ( r_- r_+ {\rho}
  + {\rho} r_- r_+ - 2r_+{\rho} r_- )
  \\ & -[\gamma_{e}(\Omega,\Delta)/2] ( r_+ r_- {\rho} +
  {\rho} r_+ r_- - 2 r_- {\rho} r_+ ) 
  \\ & - i[r_z,{\rho}]\Delta E(\Omega,\Delta) - i[H_{\mathrm{dot}},{\rho}],     
  \end{aligned}
  \label{eq:4}
\end{equation}
where we have made explicit the time dependence of the decay rates due
to the dependence on the slow variation of the parameters $\Omega(t),
\Delta(t)$. This is a time-dependent generalization of the standard
form\cite{Wangsness1953} obtained from the secularized Born-Markov
approximation in the interaction picture, as used in some related
contexts.\cite{Stace2005,Gauger2008} In addition to now preserving
positivity, it makes explicit the origin and nature of decay terms;
the damping appears as a Lindblad form describing transitions between
the instantaneous dressed states, with the phonon absorption and
emission rates\begin{align}
  \gamma_a=2\left(\frac{\Omega}{2\Lambda}\right)^2 \pi J(\Lambda) n(\Lambda), \label{eq:absrate}\\
  \gamma_e=2\left(\frac{\Omega}{2\Lambda}\right)^2 \pi J(\Lambda)
  [n(\Lambda)+1].\label{eq:emrate}\end{align} These rates are shown in
Fig.\ \ref{fig:dampfig} for the spectral function (\ref{eq:modelspecdens}).

\begin{figure}
  \ifpdf
  \includegraphics[width=3.2in]{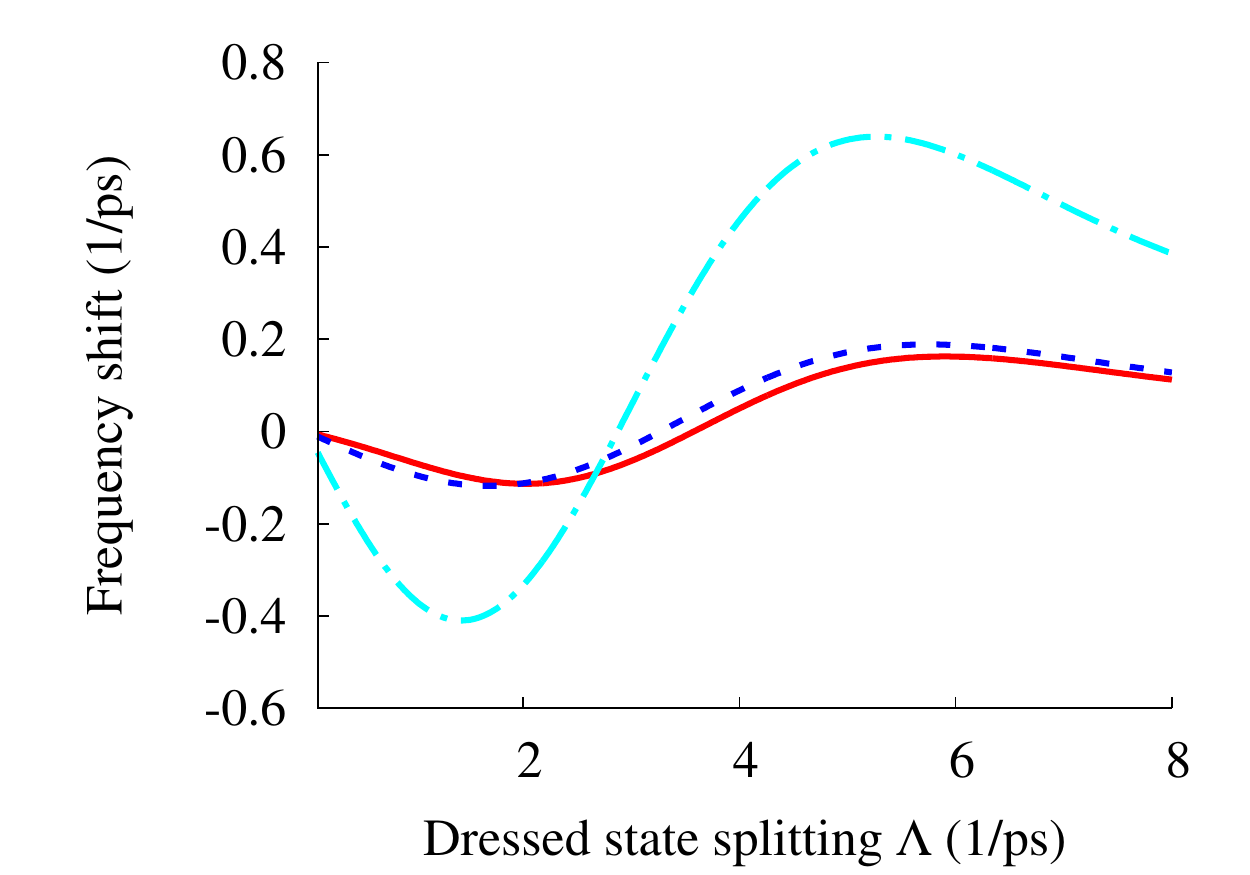}
  \else
  \includegraphics[width=3.2in]{lamb-shifts.eps}
  \fi
\caption{(color online). Phonon-induced change in the dressed-state
  splitting $\Delta E$ [see Eq.~(\ref{eq:lambshift})] divided by the
  squared ratio of Rabi splitting to dressed-state energy splitting,
  i.e., $\Delta E/(\Omega^2/\Lambda^2)$, as a function of the
  dressed-state energy splitting $\Lambda$. Curves are temperatures 1~K
  (solid), 10~K (dashed), and 50~K (dot-dashed).}
\label{fig:lamb}
\end{figure}

In addition, Eq.~(\ref{eq:4}) includes a phonon Lamb shift (see Fig.\
\ref{fig:lamb}): the energy splitting of the dressed states is now
$\Lambda+\Delta E$ with\begin{equation} \Delta
  E=-\left(\frac{\Omega}{2\Lambda}\right)^2 2\Lambda \int \frac{J(\nu)
    \coth(\nu/2k T)}{\nu^2-\Lambda^2}
  d\nu.\label{eq:lambshift}\end{equation}

Figure \ref{fig:dampfig} may be used to establish the validity of
Eq.~(\ref{eq:4}), by comparing the decay rates with the extent of
their frequency dependence. We see that at the highest temperatures
shown the peak damping rates, and therefore the linewidths, can become
a significant fraction of the width of the spectral function, as
illustrated in Fig.~\ref{fig:timescales}. In this regime, the Markovian
approximation that $\tilde\rho(t)$ varies slowly on the time scale
$1/\omega_c$ breaks down, and a quantitative analysis requires the
solution of a non-local equation.\cite{Luker2012,Reiter2012} However,
as can be seen from the linewidths in Fig.\ \ref{fig:timescales}, we
expect qualitatively reasonable results over much of the parameter
regimes shown, and the approximations should be quantitatively
accurate in the low temperature regime, below $20\,\mathrm{K}$, typical
of most coherent control experiments. We note that the Markov
approximation amounts to approximating the spectral function with its
constant value at the dressed frequency $\Lambda$. At high
temperatures, the damping at any one splitting will sample a finite
range of the spectral function, so that we expect a weaker dependence
of the effective linewidth on the energy splittings than indicated
here, as well as non-Lorentzian emission lines.  Similarly, we expect
the Markov approximation to overestimate the Lamb shift and its
frequency dependence at high temperatures.

As well as becoming invalid at high temperatures, where the effective
linewidth becomes large due to scattering, the approximations
used above also fail if the time dependence of the parameters
$\Delta(t), \Omega(t)$ becomes too strong.  This is due to the finite
bandwidth $1/\tau_{\text{chirp}} \sim
\dot{\Delta}/\Delta$ arising from the time dependence of the
parameters.  Alternatively one may understand this as arising from the
fact that for small enough $\tau_{\text{chirp}}$, the bath correlation
time no longer is the shortest time scale in the problem.

The density matrix evolution described above is a complete description
of the time evolution of the system.  In some cases, it can be useful
to write this in an alternative representation, by considering
the time evolution of the components of the Bloch vector, which we write
here for completeness:
\begin{widetext}
\begin{equation}
  \begin{aligned}
  \label{eq:14}
  \dot{s}_x &= - \frac{\Omega}{2\Lambda}(\gamma_a - \gamma_e)
  - \left[ \frac{\Delta^2+2\Omega^2}{2\Lambda^2} (\gamma_a + \gamma_e)
    \right] s_x 
    - \Delta s_y 
    +%\\&+
    \frac{\Delta \Omega}{2\Lambda^2} (\gamma_a + \gamma_e) s_z,
    \\
    \dot{s}_y &= \Delta s_x - (\gamma_a + \gamma_e) s_y/2 + \Omega s_z,
    \\
    \dot{s}_z &=
     \frac{\Delta}{2\Lambda} (\gamma_a - \gamma_e)
    + \frac{\Delta \Omega}{2\Lambda^2} (\gamma_a + \gamma_e) s_x
    - \Omega s_y
    -%\\&-
    \left[  \frac{2\Delta^2+\Omega^2}{2\Lambda^2} (\gamma_a + \gamma_e)
      \right] s_z.
    \end{aligned}
\end{equation} Note that here we have neglected the small Lamb shift.
\end{widetext}

\section{Inversion in ARP}
\label{sec:arpinv}

\begin{figure}
  \ifpdf
  \includegraphics[width=3.3in]{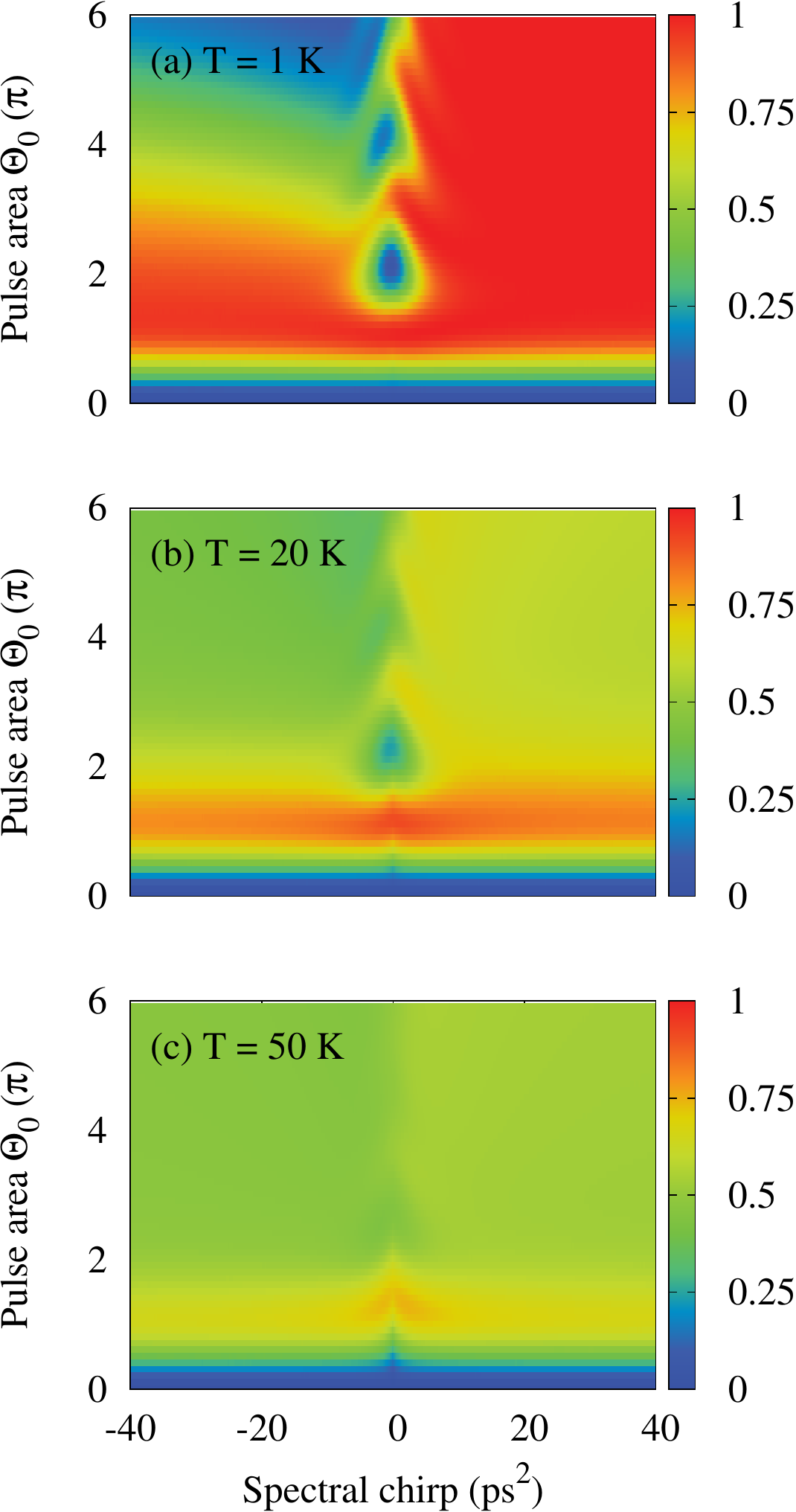}
  \else
  \includegraphics[width=3.3in,clip=true,trim=0 5.5cm 0 0]{temp1-phonons.eps}
  \fi
  \caption{(color online). Final exciton occupation probability
    following an ARP pulse of the form given in
    Eq.~(\ref{eq:kuhn-pulse}), calculated using the time-dependent
    Lindblad form given in Eq.~(\ref{eq:4}).  Calculated using
    $\tau_0=2$ ps, at temperatures 1~K (a), 20~K (b) and 50~K (c), for
    direct comparison with Fig. 2 of Ref.~\onlinecite{Luker2012}.  For
    low temperatures, and slow sweeps, the results are very similar;
    for larger temperatures and higher chirp rates the limited Markov
    approximation used in deriving Eq.~(\ref{eq:4}) becomes invalid.}
  \label{fig:gaussian-arp}
\end{figure}

The above results apply in general to any time-dependent pulse
sequence.  We can in particular consider pulses corresponding to
adiabatic rapid passage, in which the detuning $\Delta(t)$ is swept
smoothly through zero with the intention of creating a one-exciton
state. Under such a pulse, there is an avoided crossing between the
zero- and one- exciton states of Eq.~(\ref{eq:ham}), generated by the
driving field. We aim to adiabatically follow a single energy level,
thus evolving from the initial ground state to the one exciton
state. Acoustic phonon effects in this process have already been
explicitly considered in Refs.~\onlinecite{Luker2012,Reiter2012}, for
fixed-bandwidth pulses of the form
\begin{equation}
  \begin{aligned}
  \Delta(t) &= - \frac{a t}{(a^2 + \tau_0^4)},
  \\
  \Omega(t) &= \frac{\Theta_0}{\sqrt{2 \pi \sqrt{a^2 + \tau_0^4}}}
  \exp\left( - \frac{t^2 \tau_0^2}{2 (a^2 + \tau_0^4)} \right),
  \end{aligned}
  \label{eq:kuhn-pulse}
\end{equation} where $\Theta_0$ is the area of the bandwidth-limited pulse before the chirp is applied, and $a$ is the spectral chirp.\cite{Malinovsky2001} The results obtained from Eq.~(\ref{eq:4}) for this form of pulse,
plotted on a similar scale to those in Fig. 2 of
Ref.~\onlinecite{Luker2012}, are shown in
Fig.~\ref{fig:gaussian-arp}. As expected from the discussion above,
the results are very similar at low temperatures and slow sweep rates,
where the conditions for the Born-Markov approximation are well
satisfied.

As discussed in Ref.~\onlinecite{Luker2012}, the asymmetry about the
line $a=0$ at low temperatures arises because absorption
processes can be neglected $\gamma_a \ll \gamma_e$ for $T \ll
\omega_c$, and as is clear from Eq.~(\ref{eq:4}), emission can only
occur when the dot is in the higher-energy dressed state. For
$a<0$ the ARP protocol attempts to follow this higher energy
dressed state, so that the process is susceptible to phonon emission,
whereas for $a>0$ it is not. The simplicity of Eq.~(\ref{eq:4})
allows one to further see that for the range plotted, the values of
$\Lambda(t=0) = \Omega(t=0)$ lie below the peak of scattering rates
(see Fig.~\ref{fig:dampfig}), hence the decrease of inversion with
increasing pulse area visible within Fig.~\ref{fig:gaussian-arp}. For
larger pulse areas, the central value of $\Lambda$ can exceed this
peak (at $\Theta_0\approx 6\pi$ for $a=0$), and inversion then
increases with pulse area. 

\subsection{Optimization of ARP}
\label{sec:optimisation-arp}

The relatively lightweight effort of simulating Eq.~(\ref{eq:4})
allows one to rapidly investigate the effects of other potential pulse
shapes and ARP protocols beyond that in Eq.~(\ref{eq:kuhn-pulse}).  In
the absence of decay, the question of how the final excited state
population can be optimized for a given pulse area has been
extensively studied by \citet{Guerin2002}.  By considering the leading
order non-adiabatic effects, they showed that these were minimized in
the case where $\Lambda(t)$ was independent of time.  This implies
that in the limit of large pulse areas (deep in the adiabatic regime),
maximum excitation should be reached when the chirp rate is adjusted
to match this condition.

\begin{figure}[htpb]
 \ifpdf
  \includegraphics[width=3.3in]{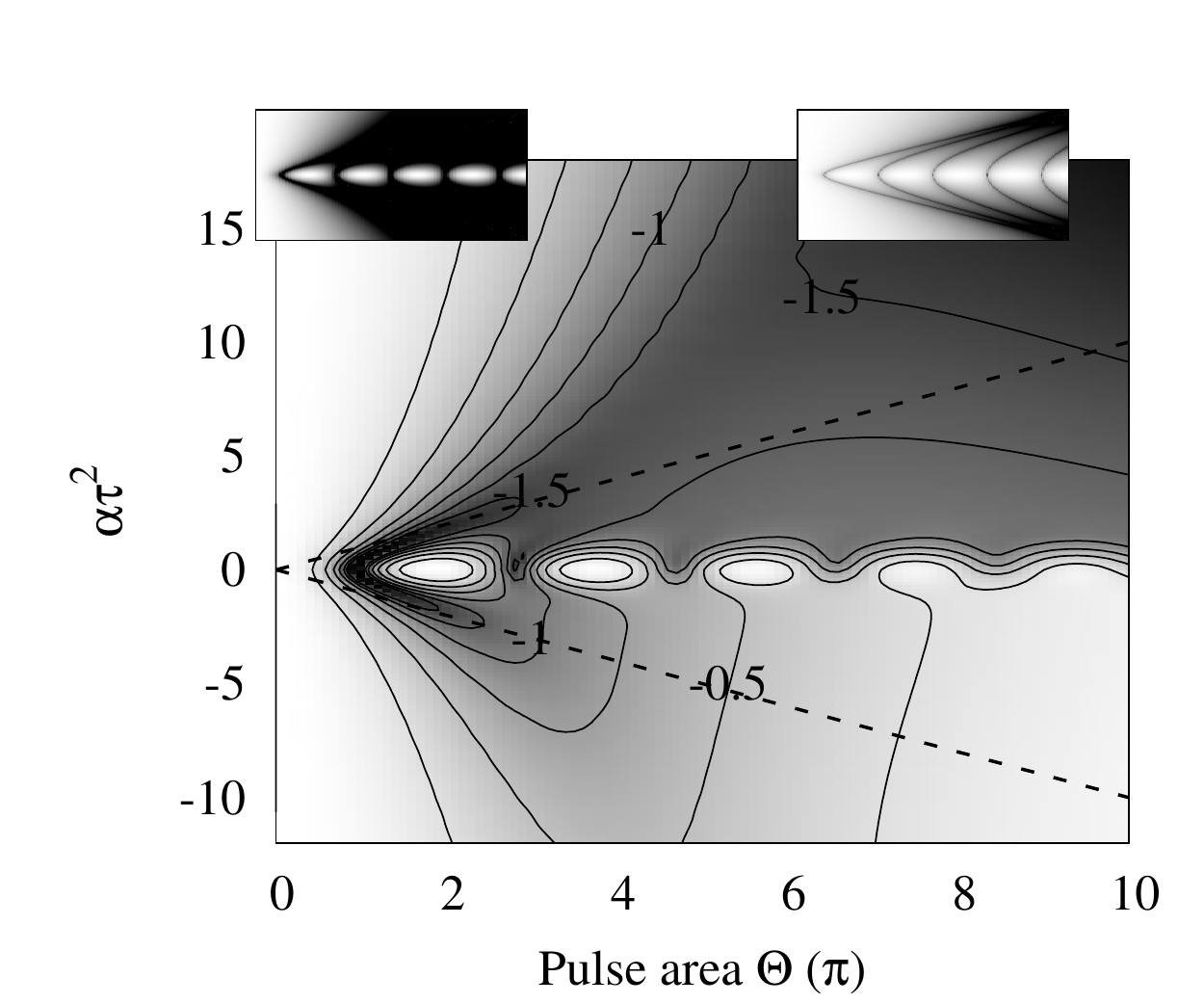}
  \else
  \includegraphics[width=3.3in]{arp-contours.eps}
  \fi
  \caption{Final exciton occupation probability for the sech pulse
    [Eq.~(\ref{eq:5})] with $\Omega_0=\Theta/\tau, \Delta_0=\alpha
    \tau$ for $\tau = 6$ ps at a temperature $4$ K. The gray scale and
    contour labels show $\log_{10} (1-P)$ where $P$ is the final
    occupation probability. Insets show the case without phonon
    induced dephasing, as in Ref.~\onlinecite{Guerin2002}, on the same
    gray scale (from -2 to 0, left inset), and on a gray scale with a
    larger range (from -10 to 0, right inset).  Main panel shows the
    effects of dynamical dephasing on the same pulse shape. Dashed
    line indicates the condition $\Omega_0=\Delta_0$ which gives the
    optimal transfer in the absence of dephasing. }
  \label{fig:sech-pulse}
\end{figure}

The time dependence of $\Delta, \Omega$ given in
Eq.~(\ref{eq:kuhn-pulse}) cannot achieve a time independent $\Lambda$.
Instead, other pulse shapes need to be considered, such as:
\begin{equation}
  \label{eq:5}
  \Delta(t) = -\Delta_0 \tanh\left(\frac{t}{\tau}\right), \qquad
  \Omega(t) = \Omega_0 \sech\left(\frac{t}{\tau}\right),
\end{equation} which have a pulse area $\Theta$ when $\Omega_0=\Theta/\tau$.
As discussed by \citet{Guerin2002}, the optimum condition $\Omega_0 =
\Delta_0$ arises from the convergence of lines originating from
the maxima of the Rabi oscillations.  This is shown in the inset of
Fig.~\ref{fig:sech-pulse}. However, although such pulses are optimal
in the isolated case, the differences in excitation near this line are
exponentially small, and entirely dwarfed by the effects of phonon
induced dephasing.  In the presence of dephasing, not only are the
sharp, exponentially small features washed out, but the optimum chirp
rate $\Delta_0/\tau$ moves to significantly larger values, due to the
reduction of the dephasing rate at large $\Lambda$.

\subsection{Thermalization enhanced inversion}
\label{sec:therm-enhanc-invers}

The discussion of the effects of acoustic phonons so far has been in
terms of their reducing the final state inversion as compared to
near-perfect inversion achieved deep in the adiabatic regime.  There
exists however a significant range of experimental conditions for
which coupling to phonons can instead enhance the final state
inversion.  This effect has recently been discussed by
\citet{Reiter2012} in the context of compensating for detuning of
quantum dots.  Even in the absence of detuning, coupling to phonons
can enhance the final state inversion. An increase in the efficiency
of adiabatic transfer due to damping processes has also recently been
reported for a many-boson model.~\cite{Wen2012}

\begin{figure}[htpb]
  \centering
  \ifpdf
  \includegraphics[width=3.2in]{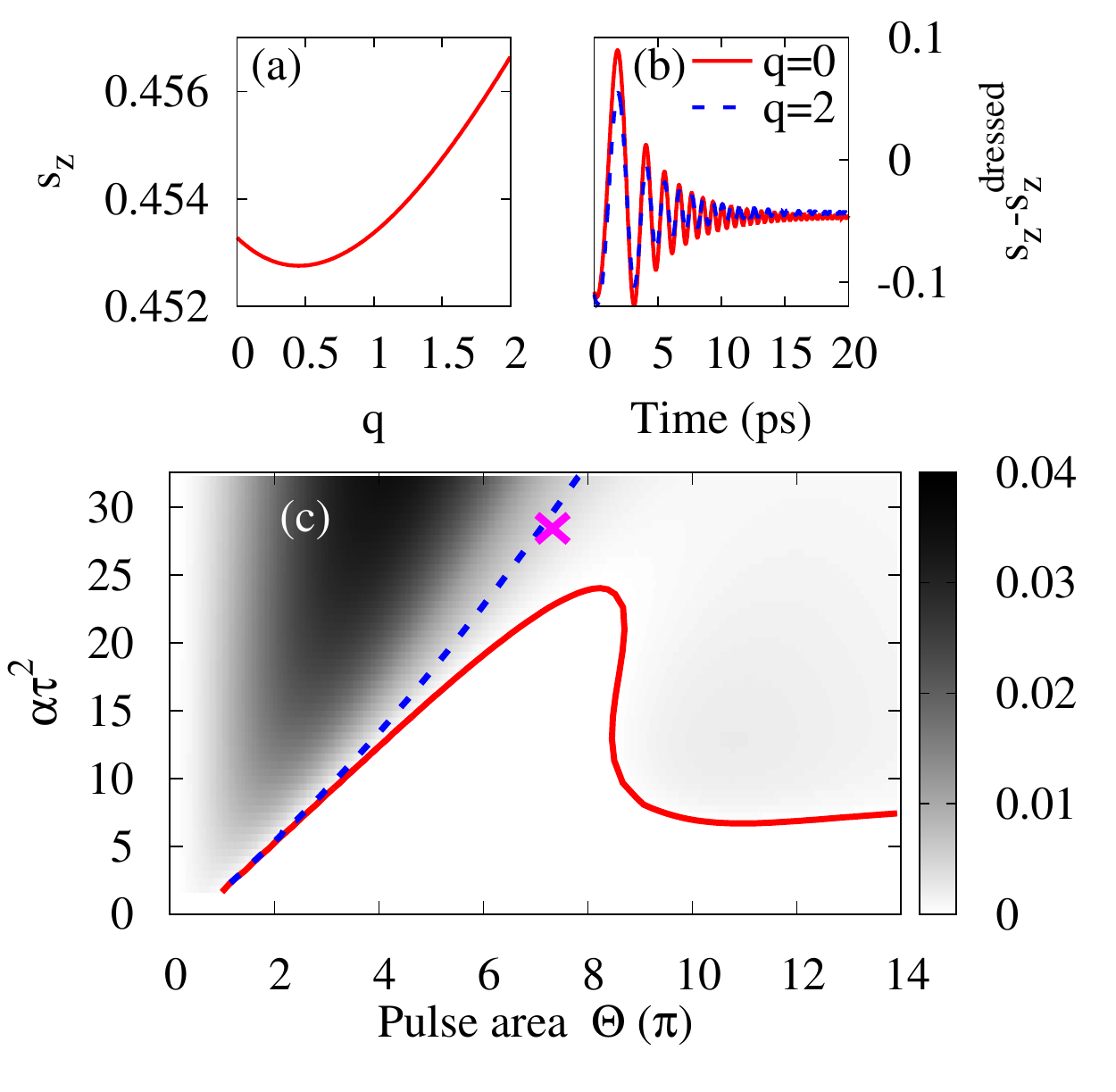}
  \else
  \includegraphics[width=3.2in]{inversion-vs-q.eps}
  \fi
  \caption{(color online). Dependence of inversion on strength of phonon coupling in
    ARP.  Panel (a) indicates how for a specific value of $\alpha,
    \Omega_0$, there is a non-monotonic dependence of inversion upon
    phonon coupling. Panel (b) indicates the time evolution of $s_z$
    at two values of q, relative to its ideal value for adiabatically
    following the dressed states. Panel (c) maps where non-monotonic
    dependence on q arises.  This is determined by two quantities: the
    grayscale indicates the value of $\text{Max}_{0<q<2}(d s_z/d q)$,
    and the solid red line is the boundary where this is zero.  The
    blue dashed line indicates where $d s_z/dq|_{q=0}=0$.  Between
    these lines, non-monotonic dependence as seen in panel (a) occurs.
    The magenta cross indicates the conditions used for panels (a) and
    (b). $T=4$~K, $\tau=5.68$~ps}
  \label{fig:thermalization-enhanced}
\end{figure}

Figure~\ref{fig:thermalization-enhanced} illustrates this potential
enhancement by showing how the final inversion is affected if the
decay rates are rescaled by a factor $q$, i.e., $J(\omega) \to q
J(\omega)$, considering a pulse shape:
\begin{equation}
  \label{eq:12}
  \Omega = \Omega_0 \sech\left(\frac{t}{\tau} \right), \qquad
  \Delta  = -\alpha t.
\end{equation}
  For a wide range of
parameters $\Theta, \alpha$ the dependence on $q$ is non-monotonic:
small coupling to phonons decreases the inversion, but further
increase in coupling then increases final inversion.  This
non-monotonic behavior [shown in
Fig.~\ref{fig:thermalization-enhanced}(a)] exists throughout the
region between the solid and dashed lines in
Fig.~\ref{fig:thermalization-enhanced}(c).  For large chirp rates and
weak pulses, the inversion without coupling to phonons is already
poor, and coupling to phonons increases the inversion; this
corresponds to the behavior above the dashed line in
Fig.~\ref{fig:thermalization-enhanced}(c).

The origin of this enhancement at large $q$ can be understood by
considering that for large $q$, the quantum dot state will come to
thermal equilibrium with the phonon bath.  Since the coupling to
phonons depends on the prefactor $q (\Omega/\Lambda)^2$, the coupling
to phonons will eventually switch off as $\Omega \to 0$.  However, the
larger the value of $q$, the later this switch off occurs, and so the
longer system maintains thermal equilibrium with phonons.  As the
detuning $\Delta$ continues to increase at late times, the inversion
of this equilibrated state therefore increases with increasing $q$.

\section{Phonon Spectroscopy}
\label{sec:phononspec}

A second application of the relative simplicity of Eq.~(\ref{eq:4}) is
to see how the phonon density of states can be recovered from
spectroscopy using an appropriately designed pulse sequence.  This
would in principle allow direct experimental confirmation of the
 model phonon coupling $J(\omega) \propto \omega^3
\exp(-\omega^2/\omega_c^2)$ as widely
used\cite{Ramsay2010,Ramsay2010a,Ramsay2011} in modeling quantum
dots.  In the following we present and compare two approaches to this,
based on either short-time or long-time behavior, incorporating
spontaneous decay in the long-time process.

\subsection{Modified ARP protocol}
\label{sec:modif-arp-prot}

The short-time approach uses a modified version of an ARP-like pulse
sequence. We consider an ARP protocol divided into pieces, so that the
sweep of $\Delta(t)$ is interrupted by a wait period $T_w$ at a value
$\Delta_w$ before completing the ARP sweep.  The initial and final
sweeps serve to map between initial or final eigenstates of $s_z$ and
eigenstates of $r_z$ during the wait period.  The final inversion
depends on the effect of the phonons during the wait time, sampling
the phonon density of states at a frequency $\Lambda_w =
\sqrt{\Delta_w^2 + \Omega_w^2}$.  During this wait time,
Eq.~(\ref{eq:4}) reduces to
\begin{math}
  \dot{p}_{\downarrow} = - \gamma_a p_{\downarrow} + \gamma_e p_{\uparrow},
  \dot{p}_{\uparrow} =  \gamma_a p_{\downarrow} - \gamma_e p_{\uparrow}
\end{math}
in terms of the diagonal elements in the $\vec{r}$ basis.  The
deviation from inversion of the final state after a given wait time
$T_w$ is thus given by:
\begin{equation}
  \label{eq:6}
  \frac{1}{2}-s_z
  \simeq
  \frac{\gamma_{a,e}}{\gamma_a + \gamma_e}
  \left(1 - e^{-(\gamma_a + \gamma_e)T_{w}} \right),
\end{equation}
where $\gamma_{a,e} = \gamma_{a,e} (\Omega_w, \Delta_w)$, and the
numerator of the right hand side is $\gamma_{a}$ for a forward sweep
($\Delta$ decreasing with time) and $\gamma_{e}$ for a reverse sweep
($\Delta$ increasing with time).  For long wait times, the excitations
reach thermal equilibrium with the phonon bath, as
expected\cite{Cresser1992} and so $s_z$ becomes independent of the
phonon density of states.  For short wait times $(\gamma_a +
\gamma_e) T_w \ll 1$, one finds $\frac{1}{2}-s_z \simeq
\gamma_{a,e}(\Omega_w, \Delta_w) T_w$, thus by varying $\Omega_w$ one can
directly map out the damping rate.

\begin{figure}[htpb]
  \centering
  \ifpdf
  \includegraphics[width=3.2in]{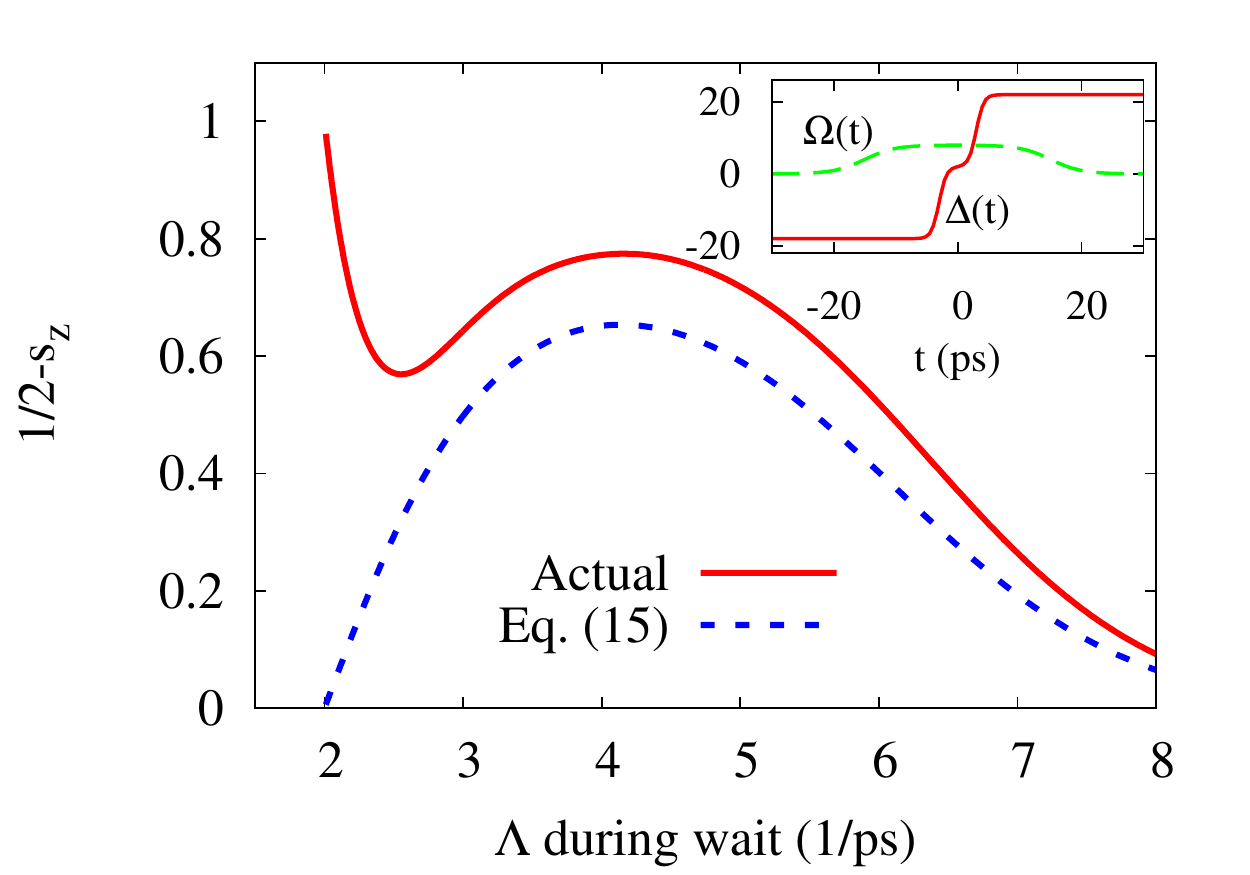}
  \else
  \includegraphics[width=3.2in]{phonon-mapping.eps}
  \fi
  \caption{(color online). Phonon spectroscopy using divided ARP
    pulse.  Main figure, deviation from inversion following the
    protocol in Eq.~(\ref{eq:7}) (solid), compared with the
    approximation of Eq.~(\ref{eq:6}) (dashed line), plotted by
    varying $\Omega_w$.  Inset, time dependence of $\Delta(t)$ (solid)
    and $\Omega(t)$ (dashed), with parameters
    $\Delta_w=2\,\text{ps}^{-1}, \Delta_0=-10\,\text{ps}^{-1},
    t_\Delta=T_w/2 = 3\,\text{ps}, \tau_\Delta =1.2\,\text{ps},
    t_\Omega=15\,\text{ps}, \tau_\Omega=5\,\text{ps}$, $T=4\,$K.}
  \label{fig:arp-spectroscopy}
\end{figure}

In order to extract the phonon density of states with some accuracy,
the pulse sequence must be carefully chosen.  At low temperatures,
$\gamma_a \ll \gamma_e$, and so spectroscopy using the forward sweep
is hard to achieve --- the reduction in inversion is tiny for times
$T_w$ such that $T_w (\gamma_a + \gamma_e) \lesssim 1$ and tends to be
dwarfed by effects of non-adiabaticity.  Using a ``reverse'' ARP pulse
produces a clearer signal.  However, in order to have the signal
dominated by the waiting time, it is necessary for the wait time to be
longer than the sweep, and the sweep to be sufficiently slow for all
values of $\Omega_w$.  For this to be compatible with $T_w \gamma_e
\lesssim 1$, it is helpful to choose $\Delta_w \ne 0$ so that the rate
$\gamma_e$ is suppressed by a factor $(\Omega_w/\Lambda_w)^2 < 1$.  Combining
these considerations, the pulse sequence:
\begin{align}
  \label{eq:7}
  \Delta(t) &= \Delta_w
  +\Delta_0 \left[
    \tanh\left(\frac{t+t_\Delta}{\tau_\Delta}\right)
    +
    \tanh\left(\frac{t-t_\Delta}{\tau_\Delta}\right)
  \right]
  \nonumber\\
  \Omega(t)&=-\frac{\Omega_w}{2}
  \left[ 
    \tanh\left(\frac{t+t_\Omega}{\tau_\Omega}\right)
    -
    \tanh\left(\frac{t-t_\Omega}{\tau_\Omega}\right)
  \right]
\end{align}
gives the results shown in Fig.~\ref{fig:arp-spectroscopy}  with
parameter values given in the figure caption.

There is a reasonable match between the prediction of Eq.~(\ref{eq:6})
for decay during the wait time and the actual inversion, but the match
is not perfect due to the effects of phonons during the initial and
final sweep, as well as some remaining degree of non-adiabaticity (at
small $\Omega_w$).  Nonetheless, one may invert Eq.~(\ref{eq:6}) in
order to extract the phonon density of states from the measured
inversion, and the result of this procedure is shown as the dot-dashed
line in the inset of Fig.~\ref{fig:sponetaneous-spectroscopy}. The
reasonable agreement confirms that for such a pulse the final
inversion contains sufficient information to extract the phonon
density-of-states; in practice one might include corrections to
Eq.~(\ref{eq:6}) by comparing an experiment directly to
Eq.~(\ref{eq:4}).

\subsection{Steady state of driven open system}
\label{sec:steady-state-open}

An alternate approach to reconstructing the phonon density of states
arises by considering the long time behavior, allowing for both
spontaneous decay as well as coupling to acoustic phonons, i.e.,
\begin{equation}
  \label{eq:13}
  \dot{\rho} \to \dot{\rho} +   (\kappa/2) ( s_+ s_- {\rho} +
  {\rho} s_+ s_- - 2 s_- {\rho} s_+ ).
\end{equation}
Equivalently, such loss modifies the equations for the Bloch vector
components as follows:
\begin{equation}
  \label{eq:16}
  \frac{d}{dt} \left(
    \begin{array}{c}
      s_x \\ s_y \\ s_z
    \end{array}
  \right)
  \to
  \frac{d}{dt} \left(
    \begin{array}{c}
      s_x \\ s_y \\ s_z
    \end{array}
  \right)
  + \kappa
  \left(
    \begin{array}{c}
      s_x/2 \\ s_y/2 \\ s_z+1/2
    \end{array}
  \right).
\end{equation}

In particular, we consider the long-time behavior under a constant
driving, $\Omega(t) = \Omega_0, \Delta(t) = \Delta_0$.  In this case,
the long-time behavior should be understood as a steady state arising
from the balance of coupling to the phonon reservoir and the decay due
to coupling to a photon reservoir.  The pumping term $\Omega$ is
necessary only in order to enable the phonon coupling to create and
destroy excitations of the quantum dot.  In the absence of spontaneous
decay terms (no photon reservoir), the steady state would trivially be
the thermal equilibrium of the system Hamiltonian, due to
equilibration with the phonon reservoir.  In this case, the phonon
density of states does not appear, and only the phonon temperature
matters.  However, the coupling to the photon reservoir drives the
system into a non-thermal steady state, in which the balance of
spontaneous decay rate and phonon coupling determines the state.  One
may then read out the phonon density of states from the steady-state
inversion achieved.

In the limit of small decays, i.e., $\kappa, \gamma_a, \gamma_e \ll
\Omega, \Delta$, one may analytically solve the equations for the
Bloch vector components, given in Eqs.~(\ref{eq:14}) and (\ref{eq:16}),
by expanding the steady-state equations with respect to the decay
rates. In the limit of vanishing decay terms $\kappa, \gamma_a,
\gamma_e \to 0$, it is clear that the steady state of
Eq.~(\ref{eq:14}) requires $\Delta s_x + \Omega s_z =0, s_y=0$.  Then,
including the decay terms to first order, one finds $s_y \sim
\mathcal{O}(\kappa, \gamma_a, \gamma_e)$, and so one may continue to
use $s_x =-(\Omega /\Delta) s_z$ as the solution of $\dot{s}_y=0$ up
to order $\mathcal{O}(\kappa^2, \gamma_a^2, \gamma_e^2)$. Using this,
the remaining two equations $\dot{s}_{x,z}=0$ can be solved by
eliminating $s_y$, to give:
\begin{equation}
  \label{eq:11}
  s_z =
  \frac{ - \Delta^2 \kappa +  \Lambda \Delta (\gamma_a - \gamma_e)}{
    (\Omega^2 + 2\Delta^2) \kappa +  2 \Lambda^2 (\gamma_a+\gamma_e)}.
\end{equation}
As anticipated, this expression involves the ratio of phonon and
spontaneous decay terms.  In the limit $\kappa \to 0$, one recovers
the thermal equilibrium result so that $s_z = (-\Delta/2\Lambda)
(\gamma_e - \gamma_a)/(\gamma_a+\gamma_e)$, in which case the phonon
density of states cancels, and no information about the phonon bath
(other than temperature) appears in $s_z$.  However for $\kappa \neq
0$, the steady state depends on the ratio of $\gamma^a/\kappa,
\gamma^e/\kappa$, and this in turn allows the phonon density of states
to be extracted from the final state.

\begin{figure}[htp<b]
  \centering
  \ifpdf
  \includegraphics[width=3.2in]{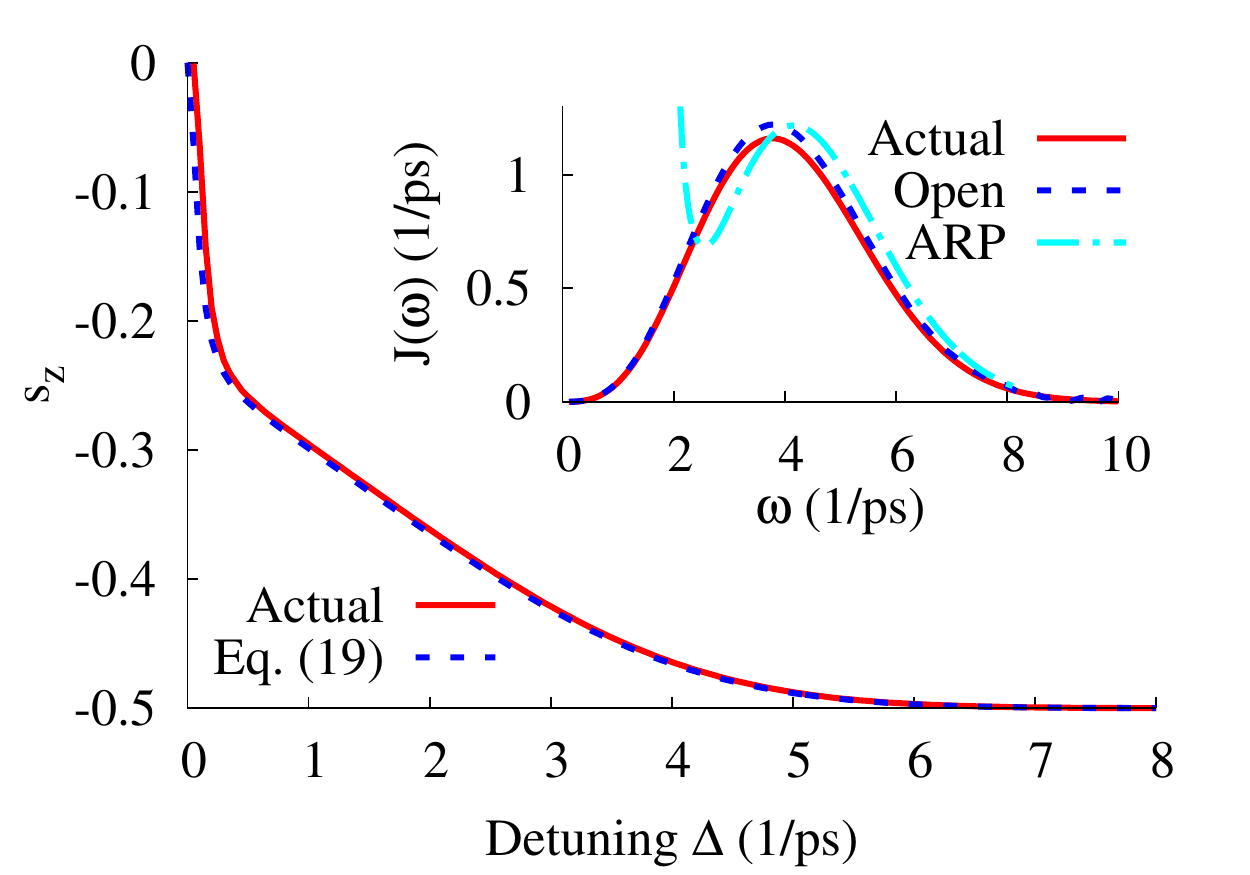}
  \else
  \includegraphics[width=3.2in]{open-mapping.eps}
  \fi
  \caption{(color online). Main figure: Steady-state value of $s_z$ with driving at
    $\Omega=0.1\,\text{ps}^{-1}$, with a spontaneous decay rate of
    $\kappa = 2\,\text{ns}^{-1}$, as a function of the detuning, at
    $T=20\,$K.  Inset: comparison of actual phonon density of states
    $J(\omega)$ with the values reconstructed by inverting
    Eq.~(\ref{eq:11}) and from the modified ARP approach, inverting
    Eq.~(\ref{eq:6}) (parameters for modified ARP as in
    Fig.~\ref{fig:arp-spectroscopy}).  }
  \label{fig:sponetaneous-spectroscopy}
\end{figure}

Figure~\ref{fig:sponetaneous-spectroscopy} shows how the phonon
density of states can be reconstructed by extracting the steady state
for a fixed value of $\Omega$, and varying the pump detuning $\Delta$.
By inverting Eq.~(\ref{eq:11}), and assuming the phonon temperature is
known, one may extract the effective phonon density of states as shown
in the inset.  The density of states reconstructed this way matches
the actual density of states used in the density matrix evolution very
closely.  It may also be possible to use recently developed
variational approaches\cite{McCutcheon2011} to extend such phonon
spectroscopy to more strongly coupled systems.

\section{Lindblad vs non-Lindblad approximations}
\label{sec:discuss}

As mentioned above, the question of which approximate Markovian
density matrix equation best corresponds to a given full density
matrix equation was discussed extensively by \citet{Dmcke1979}. The
conclusion there is that in the limit of short bath correlation times,
there exist multiple approximate Markovian equations which give have
the same order of validity, as defined by the limiting behavior of
$||\rho^{\text{approx}} - \rho^{\text{full}}||$ (where $||\ldots||$ is
the trace norm) as coupling to the bath vanishes. In other words,
there are several approximations which give the same results for the
short timescales over which perturbation theory applies. These
different approximate Markovian equations differ in regard of whether
or not one averages over rapidly oscillating terms in the interaction
picture, explicitly eliminating terms which are in any case negligible
in the limit where perturbation theory applies. Without such averaging
we reach Eq.~(\ref{eq:2}), while averaging leads instead to
Eq.~(\ref{eq:4}). However, positivity of the density matrix is only
preserved for the Lindblad form in Eq.~(\ref{eq:4}), and so
\citet{Dmcke1979} conclude that only this approach is correct.

In general, these rapidly oscillating time-dependent terms give small
changes in the density operator over small time intervals. However,
problems can arise when we use the Markov approximation to join
together many such small time intervals, and evolve the density
operator over long times. If the time-dependent terms are not treated
consistently with the Markov approximation, it might lead to an
unphysical growth of these small corrections, and potentially
unphysical results. Indeed, we note that retaining the rapidly
oscillating terms is formally inconsistent with the Markov
approximation of replacing $\tilde\rho(t^\prime)\to\tilde\rho(t)$ in
Eq.~(\ref{eq:mastereq}): this assumes that $\tilde\rho(t^\prime)$
varies more slowly than the remainder of the integrand, and in
particular more slowly than the timescale $1/\Lambda$.

\begin{figure}[htpb]
  \ifpdf
  \includegraphics[width=3.2in]{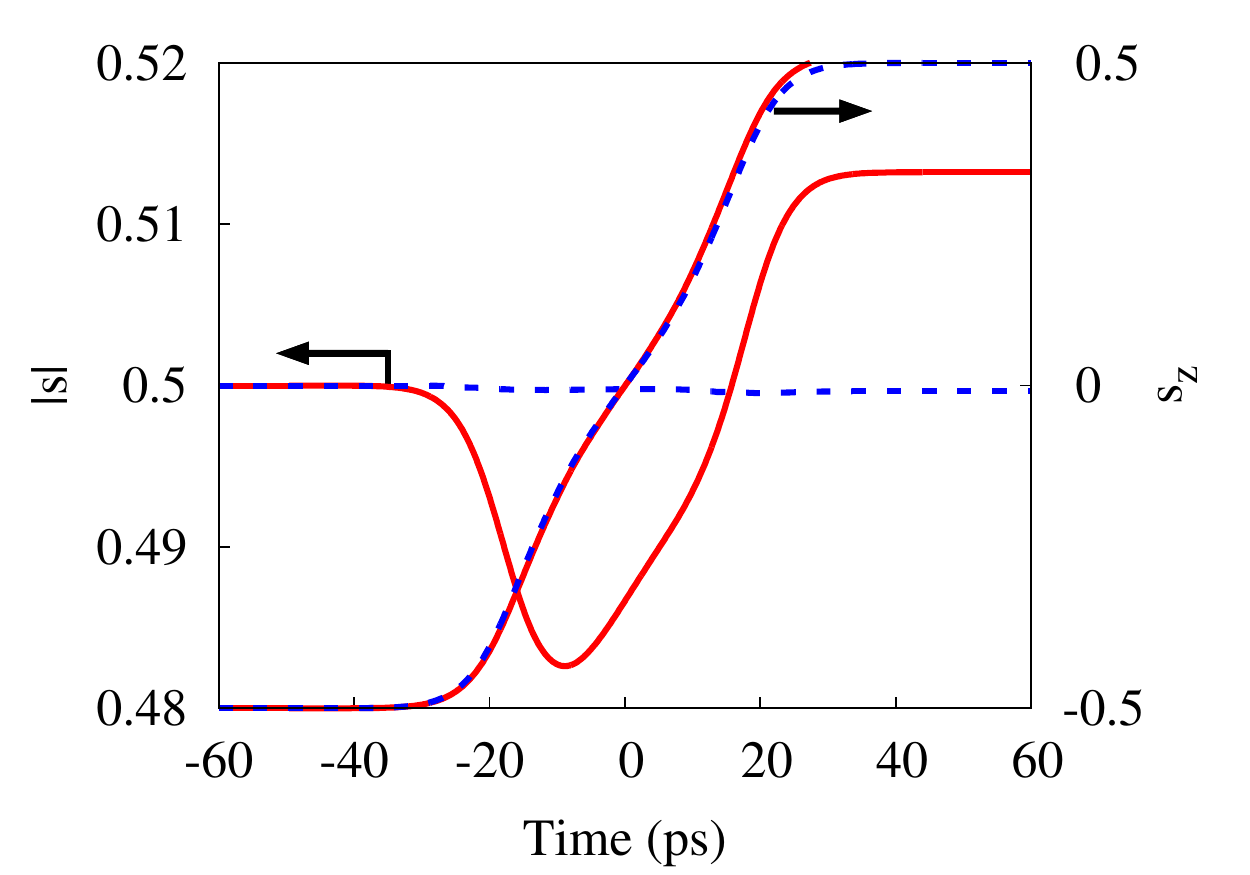}
  \else
  \includegraphics[width=3.2in]{pviolation.eps}
  \fi
  \caption{(color online). Comparison of the dynamics obtained from
    the Lindblad [Eq.~(\ref{eq:4}), dashed curves] and non-Lindblad
    [Eq.~(\ref{eq:2}), solid curves] forms of time-dependent Markovian
    approximations, for the Gaussian ARP pulse,
    Eq.~(\ref{eq:kuhn-pulse}), with $\tau_0=2$ ps, $T=1$ K,
    $a=30\,\mathrm{ps}^2$, and $\Theta_0=5\pi$. For each approach, both the
    magnitude of the pseudospin (Bloch) vector $|s|$ (left axis) and
    inversion $s_z$ (right axis) are shown. At this low temperature,
    as discussed above, the Markov approximation holds reasonably
    well, yet the non-Lindblad form leads to unphysical results
    $|s|,|s_z|>0.5$.}
  \label{fig:positivity}
\end{figure}

\begin{figure}
  \ifpdf
  \includegraphics[width=3.3in]{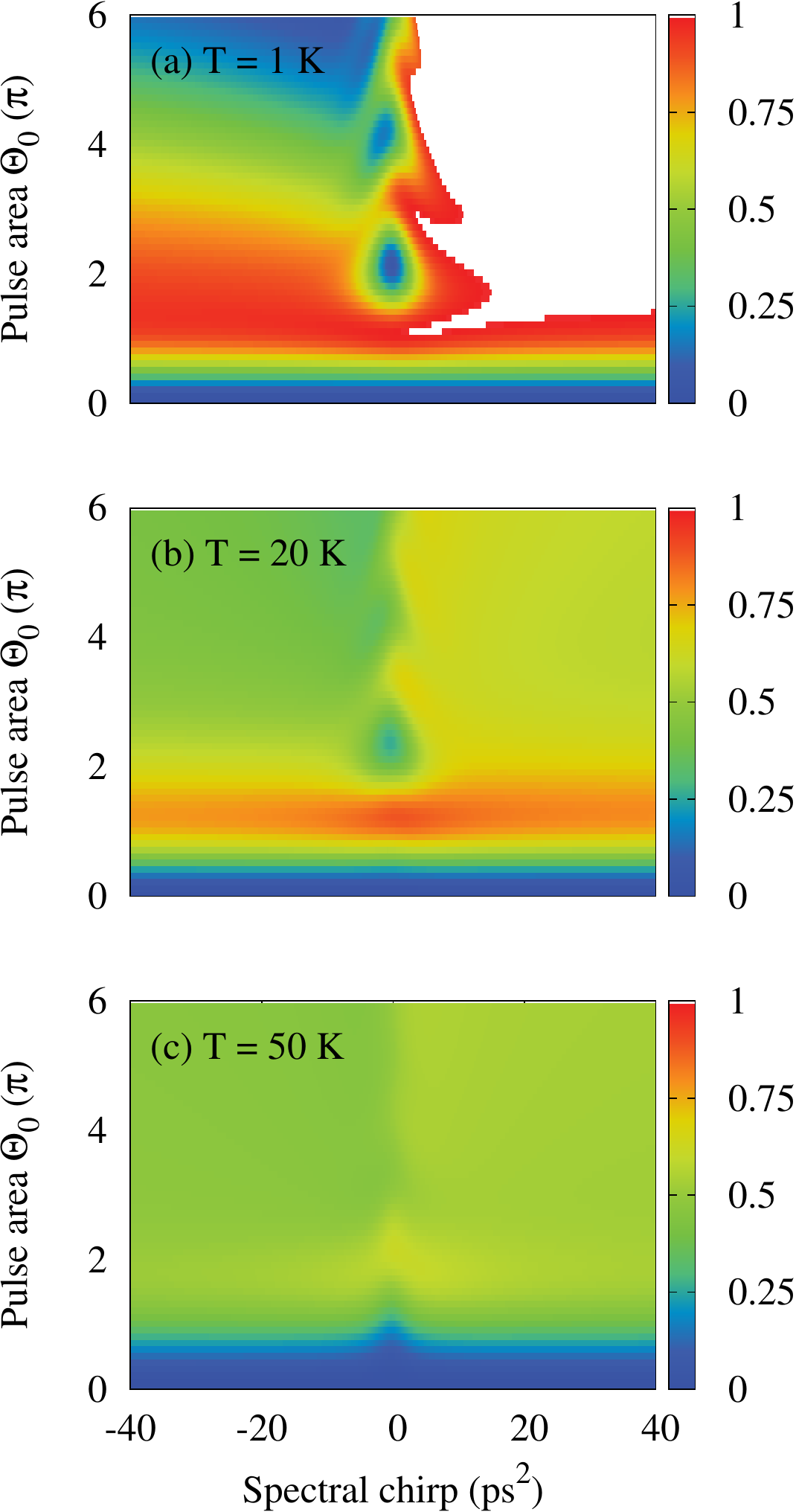}
  \else
  \includegraphics[width=3.3in,clip=true,trim=0 5.5cm 0 0]{temp1-phonons-old.eps}
  \fi
  \caption{(color online). Final exciton occupation probability
    following an ARP pulse of the form given in
    Eq.~(\ref{eq:kuhn-pulse}), calculated using the non-Lindblad form
    [Eq.~(\ref{eq:2})], for comparison with the results obtained from
    the Lindblad form [Eq.~\ref{eq:4}] shown in
    Fig.~\ref{fig:gaussian-arp}. The parameters and scale are as in
    Fig.~\ref{fig:gaussian-arp}. The results are qualitatively similar
    for many parameters. However, there is a significant region in the
    top panel (white) where the density operator obtained from
    Eq.~(\ref{eq:2}) does not remain positive, and the results become
    unphysical.  \label{fig:gaussian-arp-nonsec}}
\end{figure}

It is often assumed that such a point is irrelevant as small decay
rates (as are required for validity of the Markov approximation) imply
that any possible violation of positivity is negligible. However, for
problems involving weak decay and long time evolution, such as the
current problem, positivity violation can occur for Eq.~(\ref{eq:2})
even in regimes where the Markov approximation appears to be
valid. This can indeed lead to unphysical results, as shown for
example in Figs.~\ref{fig:positivity} and
\ref{fig:gaussian-arp-nonsec}. We note that for the Rabi oscillations
studied by \citet{Ramsay2010} no positivity violation occurs for the
parameters used and the results of Eqs.~(\ref{eq:2}) and (\ref{eq:4})
are hardly distinguishable; indeed, these equations can give similar
results for ARP pulses, as can be seen by comparing
Fig.~\ref{fig:gaussian-arp-nonsec} [which shows the results of the
non-Lindblad Eq.~(\ref{eq:2})] to Fig.~\ref{fig:gaussian-arp} [which
shows the results of the Lindblad Eq.~(\ref{eq:4})].  Despite this
similarity for much of the parameter range, Eq.~(\ref{eq:4}) is
required to produce valid results across the problems considered in
the current paper.

There have been suggestions that time-dependent density-matrix
equations which are not of the Lindblad form, such as those derived
from the time-convolutionless (TCL) approach,\cite{Breuer2007} may in
some cases ensure positivity.\cite{Whitney2008} However,
Eq.~(\ref{eq:2}) is essentially the result of the time-convolutionless
approach at second order (TCL2). The only difference is that for TCL2,
the lower limit on integrals over $t^\prime$ such as Eq.~(\ref{eq:10})
is $t^\prime=0$, rather than $t^\prime=-\infty$, corresponding to
starting the system at $t=0$ in a factorized state.  Even for time
independent $\Delta, \Omega$ this means the coefficients in the
density matrix evolution become time dependent, but eventually decay
toward a steady state value. As such, for the ARP pulse, as long as
the pulse duration is long compared to the bath correlation time,
these additional time dependencies die out, and TCL2 becomes equivalent
to the positivity-violating form in Eq.~(\ref{eq:2}).

\section{Conclusions}
\label{sec:conc}

In conclusion, we have shown how the Bloch-Redfield-Wangsness theory
may be used to derive a time-dependent Lindblad form describing the
dephasing of quantum-dot excitons by acoustic phonons in the presence
of a driving laser field. We have outlined the application of this
theory to recent ARP experiments\cite{Wu2011,Simon2011} on single
quantum dots and predict, in agreement with numerical
work,\cite{Luker2012} that phonons have a pronounced effect on ARP
even at cryogenic temperatures. Their effect can, however, be almost
eliminated by an appropriate choice of pulse shape. Furthermore, this
pulse-shape dependence could allow forms of phonon spectroscopy based
on ARP pulses or on off-resonant continuous-wave excitation. More
generally, our approach captures the physics of dynamical,
excitation-controlled dephasing, in which the driving field changes
the eigenspectrum of the dot, and hence the decoherence and scattering
rates. The resulting Lindblad form is straightforward to simulate,
gives qualitatively reasonable results over wide parameter regimes,
and is expected to be quantitatively accurate at low temperatures for
slow pulses. It can be applied to a wide variety of pulse sequences,
and the approach generalized to address a wide range of problems
relating to the decoherence of solid-state qubits, such as the
limitations on the creation of entangled states in coupled quantum
dots,\cite{Creatore2011,Unanyan2001,Kis2004,Hohenester2006,Saikin2008}
the persistence of entanglement in interacting solid-state systems,
and the emission spectra of solid-state qubits in the strong-coupling
regime.\cite{Makhlin2001,Kaer2010}

\acknowledgments{ PRE acknowledges support from Science Foundation
  Ireland (09/SIRG/I1592).  AOS acknowledges funding from the Carnegie
  Trust for the Universities of Scotland, and the St. Andrews
  Undergraduate Research Internship Program.  JK acknowledges useful
  discussions with B.~W.~Lovett, F.~B.~F.~Nissen, P.~\"Ohberg, and support
  from EPSRC (EP/G004714/2 and EP/I031014/1). 
  \vspace{-0.in} }
\appendix*

\section{Derivation of secularized density matrix equation}
\label{sec:deriv-secul-dens}

This appendix provides further details of the derivation of the
time-dependent Markovian Lindblad form in Eq.~(\ref{eq:4}).
Following the usual approach,\cite{Breuer2007} working in the
interaction picture, the effects of the system-bath coupling on
the system density matrix can be included to second order by
writing:
\begin{equation}
  \label{eq:8}
  \dot{\tilde{\rho}}
  = - \int^t_{-\infty} dt^\prime \text{Tr}_{B} \left[\tilde{H}_c(t), \left[\tilde{H}_c(t^\prime),
      \tilde{\rho}(t^\prime) \otimes \tilde{\rho}_B(0)
    \right]\right]
\end{equation}
where tilde indicates the interaction picture, $\tilde\rho_{B}$ is the
phonon bath density matrix, and $H_c$ is given in Eq.~(\ref{eq:1}).

In the case that $\Delta, \Omega$ vary slowly with time, we may
effectively transform the coupling Hamiltonian to the interaction
picture by using the instantaneous eigenstates (dressed states) giving
\begin{multline}
  H_{\mathrm{c}}(t)=P(t)\Phi(t)\\ =e^{it H_{\mathrm{dot}}} s_z e^{-it
    H_{\mathrm{dot}}} \sum_{q} (g_q b_qe^{-i\nu_q t}+g_q^\ast
  b_q^\dagger e^{i\nu q}),\label{eq:15}
\end{multline}
where $\nu_q$ is the phonon frequency. We transform to the
instantaneous dressed states, defining spin operators
$\vec{r}=R\vec{s}$, with $R$ a rotation by angle $\tan^{-1}
\Omega/\Delta$ around the y-axis. Then the coupling operators become
\begin{equation} P(t)=\frac{\Delta}{\Lambda} r_z +
  \frac{\Omega}{2\Lambda}(r_+e^{i\Lambda t}+r_- e^{-i\Lambda
    t}),
\end{equation} where $\Lambda=\sqrt{\Omega^2+\Delta^2}$ is the
splitting of the instantaneous eigenstates, and $r_{\pm}$ cause
transitions between these states. 

Note that for a time-dependent Hamiltonian, the true
interaction-picture form is obtained with a unitary transformation
involving a time-ordered exponential, $U=T e^{-i\int^t
  H_{\mathrm{dot}}(t^\prime)dt^\prime}$, and this operator is not
generally approximated by the form in Eq.~(\ref{eq:15}).  However, in
the following we use the form in Eq.~(\ref{eq:15}) only to calculate
the effects of phonons on short timescales, $t_c\sim 1/\omega_c$, with
the final equation for the density matrix obtained by undoing this
formal transformation. Thus we expect that the Hamiltonian part of the
dynamics is not approximated in the result, while the dissipative part
is correct provided $t_cd\Delta/dt,t_cd\Omega/dt<<\Lambda$. These
conditions are well satisfied for the protocols considered in this
paper.

%Formally transforming
%as in Eq.~(\ref{eq:15}) leads to terms in the Hamiltonian such as
%$td\Delta/dt, td\Omega/dt$, which are not small if there is any time
%dependence in $H$. 
%In general, the transformation to the interaction picture should lead
%to the appearance of terms in the Hamiltonian due to $d \Delta/d t$ and
%$d \Omega / dt$.  These capture the effects of non-adiabaticity
%inducing transitions between states.  Since our final expressions will
%be written back in the original Schr\"odinger picture, these effects
%will be included as regards direct effects of non-adiabaticity.  In
%principal, such non-adiabaticity can also affect the coupling to the
%baths.  For the protocols considered in this paper, such an effect is
%however small and so we neglect it.

With these forms of $P(t), \Phi(t)$, we may then follow the normal
steps of tracing over the phonon bath to give  the system
density matrix equation:
\begin{align} \dot{\tilde{\rho}}=
  - \int\! d\nu J(\nu) \int^t\!\!\!  dt^\prime \Bigl\{\Bigr.
  \left[P(t) P(t^\prime) \tilde{\rho}(t^\prime) - P(t^\prime) \tilde{\rho}(t^\prime) P(t) \right]
  \nonumber \\ \times 
  \left[(n_\nu +1) e^{-i \nu (t-t^\prime)} + n_\nu e^{i \nu (t-t^\prime)} \right]
  \nonumber\\
  -
  \left[P(t) \tilde{\rho}(t^\prime) P(t^\prime)  - \tilde{\rho}(t^\prime) P(t^\prime)  P(t) \right] 
  \nonumber \\ \times
  \left[(n_\nu +1) e^{i \nu (t-t^\prime)} + n_\nu e^{-i \nu (t-t^\prime)} \right]
  \Bigl.\Bigr\}, \label{eq:mastereq}
\end{align} 
where $n_\nu$ is the thermal occupation of the phonons at frequency
$\nu$. After performing the integrations over frequency, the remaining
integral contains factors which are sharply peaked functions of
$t-t^\prime$, decaying over a timescale $1/\omega_c$. $\tilde\rho(t')$, $\Delta(t')$, and $\Omega(t')$ vary little over this time scale, and so may be
approximated by their values at $t$. However, $P(t^\prime)$ may vary
over this time scale due to the time-dependence arising from the
transformation to the interaction picture.  If we approximate
$\tilde\rho(t^\prime) \simeq \tilde\rho(t)$ and perform no other steps, this leads
to Eq.~(\ref{eq:2}), which as noted before, is not of Lindblad form.

Following \citet{Dmcke1979}, the corresponding Lindblad form arises by
``secularizing'' Eq.~(\ref{eq:mastereq}).  This corresponds to averaging the
above equation over a time short compared to decay rates, but long
compared to the timescales of the system Hamiltonian --- the fact that
such a timescale exists is implicit in the use of a perturbative
(Born) approximation.  We start from Eq.~(\ref{eq:2}), with $P(t)$ as
defined in Eq.~(\ref{eq:3}), and writing $Q(t)$ in the generic form:
\begin{equation}
  \label{eq:9}
  Q(t) =  \Gamma_z r_z + 
  \Gamma_+ r_+ e^{i \Lambda t} +
  \Gamma_- r_- e^{-i \Lambda t}. 
\end{equation}
This follows directly from performing the integrals in
Eq.~(\ref{eq:10}), and so $\Gamma_z, \Gamma_\pm$ are various frequency
integrals over $J(\nu)$.  Multiplying $P(t), Q(t)$ and integrating
over a time long compared to $1/\Lambda$, only those terms with equal
and opposite $t$ dependence will survive, i.e. those terms involving
$r_z r_z, r_+ r_-$ or $r_- r_+$.  The secularized Eq.~(\ref{eq:2}) thus becomes:
\begin{align*}
  \dot{\tilde{\rho}}
  &=  \frac{\Delta}{\Lambda} (\Gamma_z+\Gamma_z^\ast)
  \left( r_z \tilde{\rho} r_z -  \tilde{\rho} \right)
  \\&-
  \frac{\Omega}{2\Lambda} \left[\Gamma_+
  \left(
    r_- r_+ \tilde{\rho} - r_+ \tilde{\rho} r_-
  \right)
  +
   \Gamma_+^\ast
  \left(
     \tilde{\rho}  r_- r_+  - r_+ \tilde{\rho} r_-
  \right)
  \right]
  \\&-
  \frac{\Omega}{2\Lambda} \left[\Gamma_-
  \left(
     r_+ r_- \tilde{\rho} - r_- \tilde{\rho} r_+
  \right)
  +
   \Gamma_-^\ast
  \left(
     \tilde{\rho}  r_+ r_- - r_- \tilde{\rho} r_+
  \right)
  \right].
\end{align*}
The vanishing of the phonon density of states as $\omega \to 0$
ensures that $\Gamma_z + \Gamma_z^\ast = 0$, and the remaining terms
take the form of the Lindblad decay and phonon Lamb shift terms as
given in Eq.~(\ref{eq:4}).  

%\bibliography{state-prep}


\end{document}